\documentclass[%
reprint,
superscriptaddress,
 amsmath,amssymb,
 aps,
prb,citeautoscript]{revtex4-1}

\usepackage[T1]{fontenc} 

\usepackage{graphicx}
\usepackage{dcolumn}
\usepackage{bm}
\usepackage{lipsum}
\usepackage{color}
\usepackage[normalem]{ulem}
\usepackage{simplewick}
\usepackage{qcircuit}
\usepackage{braket}
\usepackage{multirow}
\usepackage{tikz}
\usepackage{xr-hyper}
\usepackage[colorlinks=true,urlcolor=blue,citecolor=blue,linkcolor=blue]{hyperref}
\usepackage{float}
\usepackage[thinlines]{easytable}
\usepackage{booktabs,graphics}
\usepackage{array}
\usepackage{geometry}
\usepackage{chemformula}
\begin{document}

\title{Quantum algorithm for alchemical optimization in material design}

\author{Panagiotis Kl. Barkoutsos}
\affiliation{IBM Research GmbH, Zurich Research Laboratory, R\"uschlikon, Switzerland}

\author{Fotios Gkritsis}
\affiliation{IBM Research GmbH, Zurich Research Laboratory, R\"uschlikon, Switzerland}
\affiliation{King's College London, London, UK}

\author{Pauline J. Ollitrault}
\affiliation{IBM Research GmbH, Zurich Research Laboratory, R\"uschlikon, Switzerland}
\affiliation{Laboratory of Physical Chemistry,
ETH Z\"urich, 8093 Z\"urich, Switzerland}

\author{Igor O. Sokolov}
\affiliation{IBM Research GmbH, Zurich Research Laboratory, R\"uschlikon, Switzerland}

\author{Stefan Woerner}
\affiliation{IBM Research GmbH, Zurich Research Laboratory, R\"uschlikon, Switzerland}

\author{Ivano Tavernelli}
\email{ita@zurich.ibm.com}
\affiliation{IBM Research GmbH, Zurich Research Laboratory, R\"uschlikon, Switzerland}

\date{\today}

\pacs{Valid PACS appear here}

\begin{abstract}
The development of tailored materials for specific applications is an active field of research in chemistry, material science and drug discovery.
The number of possible molecules that can be obtained from a set of atomic species grow exponentially with the size of the system, limiting the efficiency of classical sampling algorithms. 
On the other hand, quantum computers can provide an efficient solution to the sampling of the chemical compound space for the optimization of a given molecular property.  
In this work we propose a quantum algorithm for addressing the material design problem with a favourable scaling.
The core of this approach is the representation of the space of candidate structures as a linear superposition of all possible atomic compositions. The corresponding `alchemical' Hamiltonian drives then the optimization in both the atomic and electronic spaces leading to the selection of the best fitting molecule, which optimizes a given property of the system, e.g., the interaction with an external potential in drug design.
The quantum advantage resides in the efficient calculation of the electronic structure properties together with the sampling of the exponentially large chemical compound space.
We demonstrate both in simulations and in IBM Quantum hardware the efficiency of our scheme and highlight the results in a few test cases. These preliminary results can serve as a basis for the development of further material design quantum algorithms for near-term quantum computers.

\end{abstract}

\maketitle

\paragraph*{Introduction.}
The chemical compound space (CCS), that is, the ensemble of possible molecules that can be constructed with a given set of atoms, is known to grow exponentially with the size of the molecular systems of interest. 
For example, in a 2004 Nature Insight issue the number of small organic molecules expected to be stable has been estimated to exceed $10^{60}$~\cite{Kirkpatrick_chemical_2004,Lipinski_navigating_2004,Dobson_chemical_2004}.
By contrast, current records held by the Chemical Abstract Services of the American Chemical Society account for only ~100 million compounds characterized so far.

The vastness of the CCS offers a formidable opportunity for the discovery of new materials, but at the same time, it poses enormous challenges. 
In fact, while the exponentially large set of possible chemical species enables the potential design of novel molecules and materials with improved properties for  numerous applications in physics, chemistry and biology, current experimental and computational techniques are still unable to perform an efficient optimization in such high-dimensional space.

Classical approaches to molecular design are currently based either on deterministic algorithms to exploit function-structure relationships using first-principle (or force-fields based) solutions of the underlying physical models (e.g., Schr\"odinger's or Newton's equations of motion) or on machine learning (ML) and regression models, like the quantitative structure-activity relationship (QSAR) techniques~\cite{Roy_a_primer_2015}.
Rational design based on the quantum mechanical framework is crucial for the unbiased exploration of CCS since it enables, at least in principle, the exact and deterministic evaluation of system properties through the calculation of expectation values. However, the computational cost associated to this approach hampers a systematic exploration of the complete CCS, limiting drastically its applicability. 
On the other hand, despite a long tradition of ML methods in pharmaceutical applications~\cite{Kubinyi_3D_1994,Faulon_the_signature_2003, Golbraikh_rational_2003, Ivanciuc_QSAR_2000, Baldi_Editorial_2011}
and many successful applications as filters applied to large molecular libraries
~\cite{Drwal2013}, the overall usefulness of ML for molecular design is still controversial~\cite{Wagner_the_evaluation_1981, Schneider2010Virtual}. 
Of different nature are the more recent ML techniques trained across the CCS and used to predict, among others, 
reorganization energies~\cite{Misra_toward_2011}, chemical reactivity~\cite{Kayala_reactionpredictor_2012}, 
and crystal properties~\cite{Schutt_how_to_represent_2014,  Meredig_combinatorial_2014}. 
The automatic generation of ML models for classical and quantum observables  
has only recently been accomplished within the rigorous realm of physical chemistry~\cite{Rupp_fast_2012}. However, even though promising, these methods are still in their infancy and therefore not yet of general applicability~\cite{Rupp_fast_2012}.

In the case of drug discovery, the main goal is often to find the `best' molecular structure that is capable to produce favourable interaction with a given biological target like, for instance, the binding pocket of an enzyme.
Also in this case, the number of accessible stable molecules that can potentially lead to a favourable drug-target interaction is immense. 
In the case of the optimization of the ligands associated to a known molecular motif (a molecular scaffold), the number of possible configurations obtained by associating a given ligand - selected from a ligand species database with $n_s$ elements - to each of the $n_p$ insertion points of a given molecular scaffold grows exponentially as $n_s^{n_p}$. 

In this work, we introduce a quantum algorithm that 
enables the efficient \textit{simultaneous} optimization of the atomic composition and corresponding electronic structure for an exponentially large set of molecules loaded as a
linear superposition of structures in the Hilbert space of a $N$-qubit quantum register.
Within this linear combination of all possible drug candidates the quantum optimization algorithm will then select a small subset of stable candidates with a favourable interaction with the external potential.
The quantum advantage of this approach is therefore twofold. On one side, we benefit from the favorable $\mathcal{O}(N^4)$ scaling for the solution of the Schr\"odinger equation (SE) in a quantum computer, and - on the other side - we can exploit the size of the qubit Hilbert space to efficiently scan the properties of an exponential set of potential drugs.

In this perspective, our quantum algorithm falls in the category of `inverse design'~\cite{Wang_designing_2006}, namely the optimization of molecular structures given a desired target property. 
As such, this approach is not limited to the design of new drugs that minimize a given ligand-receptor interaction, but it can  be easily generalized to the optimization of different properties of interest in chemistry and physics like, for instance, the optical absorption and emission of chromophores, the efficiency of new catalysts, and the prediction of binary alloys. 
This work is focusing exclusively on those aspects of the design process that determine the exponential cost of simulations and that can be addressed using a quantum computing algorithm. Additional steps including the molecular relaxation of the system and of its environment can be add without altering the scaling properties.

\paragraph*{Methods.}
In this work, we construct an `alchemical' Hamiltonian that describes a linear superposition of all possible molecular structures generated by the insertion of $n_i$ molecular fragments chosen from a set with $n_l$ elements into a molecular scaffold of a defined structure. 
As an example, one can think about a simple drug scaffold such a cholesterol derivative and the attempt to improve its interaction with a target molecule by changing a subset of its functional groups. 
Even though our quantum algorithm can be generalized to any type of ligands, this study is limited to the single atomic `mutations' of a given drug motif. 
The chemical nature of the atomic species is encoded in the nuclear charge and effective core potentials~\cite{Hellmann1935_a_new_approximation}
(ECP) that describe the effect of the atomic core electrons that  are not treated explicitly in the solution of the corresponding electronic structure equation.
The corresponding `alchemical' Hamiltonian has the form
\begin{eqnarray}
\label{Hamilt_1st}
\nonumber H(R,\alpha) &=& K_e(r)\\ \nonumber
&+&\sum_{I=1}^{N_n} 
        \sum_{s_I=1}^{N^{\rm max}_I} 
     \alpha^I_{s_I} 
     \left( 
    \sum_{i=1}^{N_e}
    V_{en}(\tilde{Z}^{s_I}_I,r_i,R_I) + v^{I,s_I}_{\rm{ECP}}(r,R_I)
    \right)\\ 
    &+&V_{nn}(\tilde{Z}^{s},R)
    + V_{ee}(r)
\end{eqnarray}
where 
$r=\{r_1,\dots,r_{N_e}\}$ is the collective vector of the electronic coordinates,
$R=\{R_1,\dots,R_{N_n}\}$ is the collective vector of the nuclear coordinates,
$e\tilde{Z}^{s_I}_I$ are the valence charges with $\tilde{Z}^{s_I}_I=Z^{s_I}_I-N_{e}^{\rm{ECP}}(s_I)$
(where $Z^{s_I}_I$ is the atomic number of atom $I$ of species $s_I$, $e$ is the electron charge, $N_{e}^{\rm{ECP}}(s_I)$ are the number of electrons of the core),
$e\tilde{Z}^{s}$ is the collective vector of all valence charges and all possible atomic species,
$I$ and $i$ are indices for the atoms and the electrons, respectively,  
$s_I$ runs over the different chemical species associated to the atomic position $I$, $N^{S_I}_I$ (with $ N^{\rm max}_I=\max_I \{N^{S_I}\}$), 
and 
$\alpha^I_s$ are the `alchemical' weights subject to the constraint $\sum_{s_I} \alpha_{s_I}^I =1 , \, \forall I$ and $\alpha$ is the collective vector of all $\alpha^I_s$.
In Eq.~\eqref{Hamilt_1st}, 
$K_e$ is the kinetic energy of the electrons,
$V_{nn}$ is the nuclear-nuclear interaction,
$v^{I,s_I}_{\rm{ECP}}$ is the potential generated by the core electrons of atoms $I$ in its `alchemical' form $s_I$, and finally
$V_{ee}$ 
is the electron-electron interaction.
All calculations are performed in the unrestricted formalism, without fixing the total electronic spin state.
The cost function that is used to score the different potential molecular candidates is given by the binding energy in the field generated 
by a set of external charges, $q_k$ placed in positions $\tilde{R}_k$ and defined by
\begin{equation} \label{Eq_cost_function}
    \Delta E(R, \alpha, \tilde{R}, q)=
    E_C(R, \alpha, \tilde{R}, q)
    -
    E(R, \alpha)
\end{equation}
where $E(R, \alpha)$ is the vacuum expectation value of the Hamiltonian in Eq.~\eqref{Hamilt_1st} for the optimized `alchemical' state $\psi(r; R,\alpha)$ evaluated with the quantum computer, $\langle \psi(r; R,\alpha) |H(R,\alpha) |\psi(r; R,\alpha)\rangle$, and $E_C(R, \alpha, \tilde{R}, q)$ is the expectation value (ground state energy) of the system in the field of the external charges governed by the Hamiltonian
\begin{eqnarray} \label{Hamilt_1st_Charges}
    \nonumber & &H_C(R,\alpha) = H(R,\alpha) \\  &+& \sum_{k=1}^{N_c} \left( \sum_{i=1}^{N_e}  \frac{e \, q_k}{|r_i-\tilde{R}_k|} + 
    \sum_{I=1}^{N_n} \sum_{s_I=1}^{N^{\rm max}_I} \alpha^I_{s_I} \frac{e\tilde{Z}^{S_I}_I q_k}{|R_I-\tilde{R}_k|} \right)
\end{eqnarray}
with corresponding ground state wavefunction $\psi'(r; R,\alpha,\tilde{R}, q)$. The last two energy contributions are referred as $V_{eq}$ and $V_{nq}$, respectively.
Note that in Eq.~\eqref{Eq_cost_function} additional repulsion terms can be added to the cost function to account for eventual contacts between the system and its environment. Furthermore, structural relaxation can be added to the optimization procedure to prevent steric contacts between the two subsystems.

The quantum algorithm requires the transformation of the `alchemical' Hamiltonian in the second quantization framework (as a fermionic Hamiltonian~\cite{romero_strategies_2017, Barkoutsos2018_quantum, kandala_hardware-efficient_2017, omalley_scalable_2017, Sokolov2019quantum, Reiher2016}).
This needs the selection of one-electron basis functions, which is commonly assumed to be the set of molecular Hartree-Fock (HF) orbitals. 
However, this would imply the solution of the HF equations for each of the exponentially many possible structures obtained by assigning different atomic species (characterized by the `valence' atomic number $e\tilde{Z}^{s_I}_I$ with $s_I$ in the set of considered elements) at each atomic position of the molecular scaffold.
To avoid this potential pitfall, we prepare the second quantized Hamiltonian in the basis of the atomic functions, that in our case is given by the Gaussian STO-3G basis set~\cite{ Jensen2013_atomic, Hehre1969_self} (the generalization to other basis sets is straightforward). We denote the elements of this basis set with $\phi_\mu(r)$ where $\mu$ is a collective index that runs over all basis functions associated to all atomic species allowed at each atomic position. The total number of such basis functions is therefore $N^{t}_b=N_n \, N_s \, N_b$, when we assume for simplicity that at each atomic position we have the same number of possible alternative atomic species ($N_s$), each one described by the same number of atomic basis functions ($N_b^{s_I}=N_b, \, \forall s_I$).
The main drawback of this choice consists in the requirement of a larger number of qubits, $N$, for the construction of the molecular wavefunction, without, however, modifying the overall scaling of the algorithm, which remains $\mathcal{O}(N^4)$. 
The Hamiltonian in Eq.~\eqref{Hamilt_1st_Charges} becomes
\begin{eqnarray} \label{Eq_Hamiltonian_2nd}
    \nonumber H^{(2)}(R,\alpha,\tilde{R},q) &=& \sum_{\mu\nu} h_{\mu\nu}(R,\alpha,\tilde{R},q) a_{\mu}^{\dagger} a_{\nu} \\ 
    &+ & \sum_{\mu\nu\kappa\lambda} \frac{1}{2} g_{\mu\nu\kappa\lambda}(R) a_{\mu}^{\dagger} a_{\nu}^{\dagger}   a_{\kappa} a_{\lambda}  \\ 
    &+ & \nonumber V_{nn} + V_{nq} \, ,
\end{eqnarray}
where the coefficients $h_{\mu\nu}(R,\alpha,\tilde{R},q)$ are the sum of the matrix elements of the one-electron terms in Eqs.~\eqref{Hamilt_1st} and~\eqref{Hamilt_1st_Charges} (i.e., the potentials $V_{en}$, $v^{I,s_I}_{\text{ECP}}$ and $V_{eq}$), and $g_{\mu\nu\kappa\lambda}$ are the two-electron integrals of the potential $V_{ee}$, which depend only implicitly on $R,\alpha,\tilde{R}$, and $q$. 
In Eq.~\eqref{Eq_Hamiltonian_2nd} the rising and lowering operators $a_{\mu}^{\dagger}$ and $a_{\mu}$ operate in the Fock space span by the atomic basis functions $\{\phi_{\mu}(r)\}_{\mu=1}^{N_b^t}$. 

Note that the `alchemical' Hamiltonian (Eq.~\eqref{Hamilt_1st_Charges}) has the same complexity as the original electronic structure problem formulated in second quantization, with the only difference that the ground state solution is now evaluated for a superposition of structures characterized by the coefficients $\alpha^I_{S_I}$. 
Thanks to the quantization in the atomic basis, we achieve to break down the exponential cost to linear, since each atom in the molecule is contributing to the total wavefunction with a set of independent orbitals of the size $N^{S_I}_I$.

The optimization of the `alchemical' system wavefunctions $\psi(r; R)$ and $\psi'(r; R,\alpha,\tilde{R}, q)$, respectively in the absence and in the presence of the external potential generated by the point charges, is performed using the variational quantum eigensolver (VQE) algorithm~\cite{peruzzo_variational_2014}.
For an initial set of parameters $\alpha$, the circuit in Fig.~\ref{fig:fig_2}(a) evaluates a trial wavefunction for the corresponding linear superposition of molecular Hamiltonians. The wavefunction is parametrized by the single qubit rotation angles, $\theta=\{\theta_1,\dots,\theta_M\}$ (where $M$ is the total number of parametrized gates) according to the hardware-efficient Ansatz described in references~\cite{kandala_hardware-efficient_2017, Barkoutsos2018_quantum}.
In the most general case, the total number of the electrons is not fixed during the optimization, but it varies as the atomic composition of the ensemble evolves, in such a way to minimize the molecular potential energy and maximize the interaction with the environment. Additional constraints enforcing the desired charge state can be also implemented.
At each VQE iteration both parameter sets, $\{\theta, \alpha\}$, are updated in order to minimize the cost function $\Delta E(R, \alpha, \tilde{R}, q)$ in Eq.~\eqref{Eq_cost_function} for fixed values of the charges and corresponding positions ($\tilde{R}, q$). 
Note that in this application we are dealing with a modified version of the original VQE algorithm in which the optimization is extended to a set of parameters $\alpha$ that defines the cost function.
At convergence, the algorithm provides the set of optimized parameters $\{\theta_{opt},\alpha_{opt}\}$, which defines the `alchemical' state that  minimizes the interaction with the environment. 
This corresponds to a superposition of possible physical states (molecules) weighted by the coefficients $\alpha_{opt}$. 
The final step consists in the selection of the most suited atomic species to be located at the atomic site $I$ of the optimized molecule, according to the sampled VQE distributions (see the results section).
The algorithm can converge towards a pool of potential candidates with similar scoring values instead of a single structure. In this case, after imposing a threshold value, it is possible to identify a small subset of molecules that can be further analyzed.

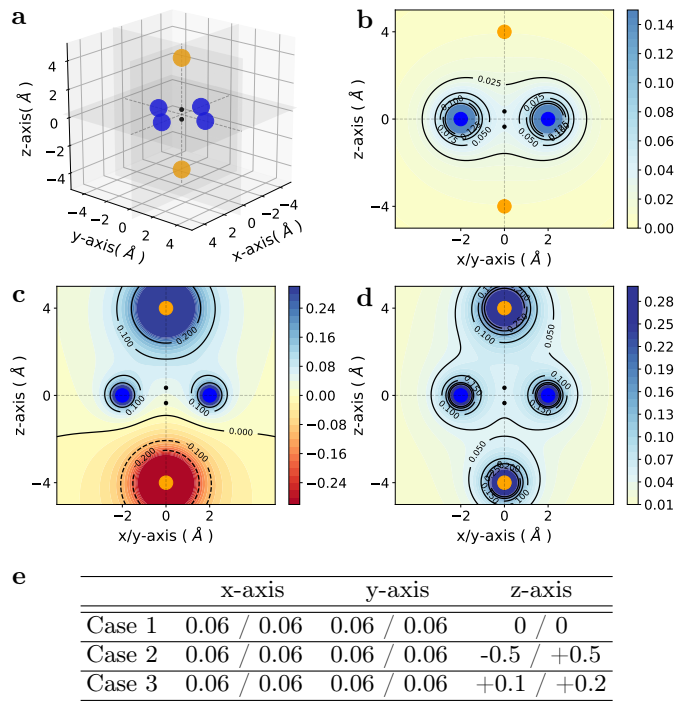
\begin{figure}[b]
    \centering
    \begin{tikzpicture}
        \node[inner sep=0pt] (russell) at (-1,0)
            {\includegraphics[width=\columnwidth]{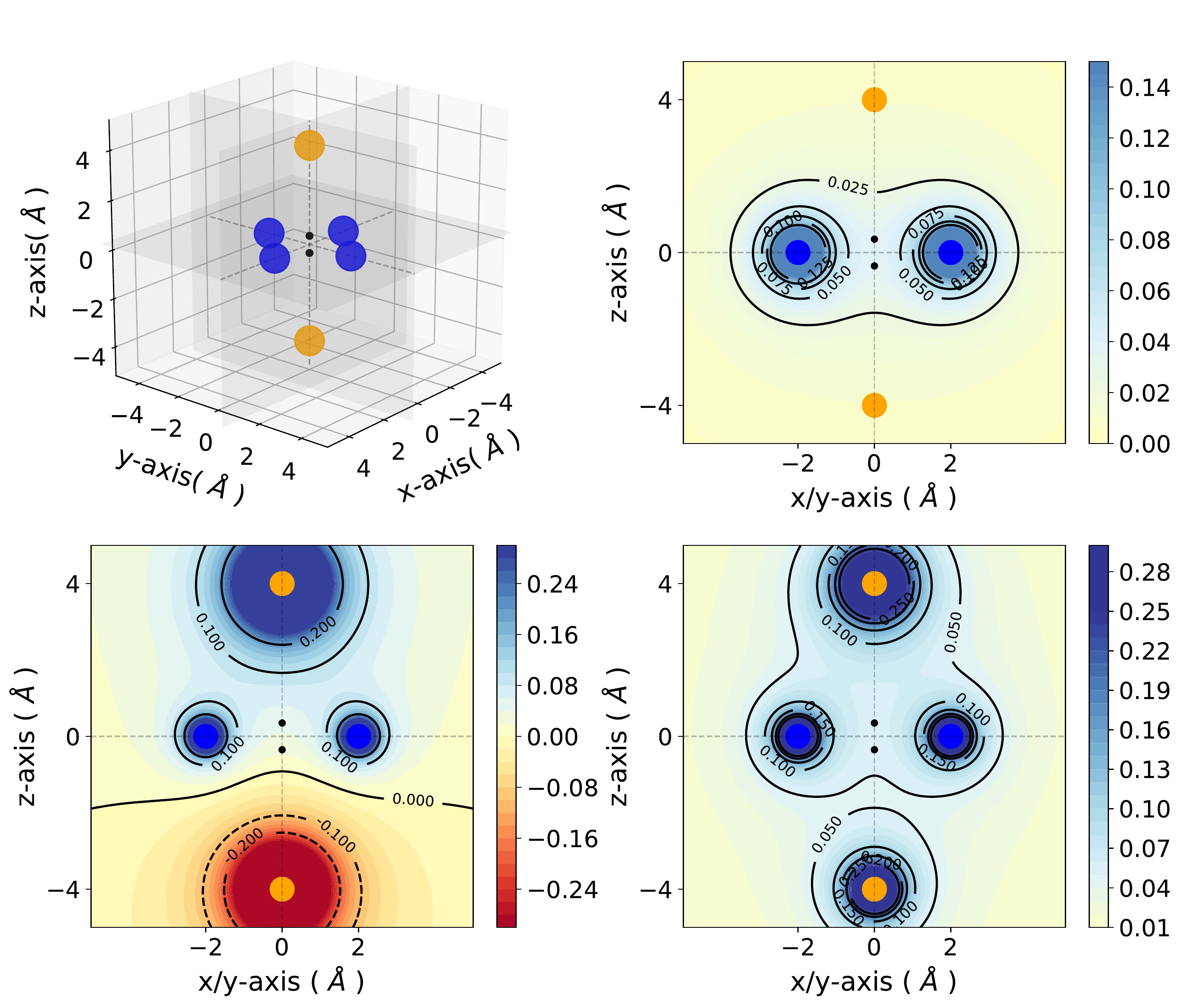}};
        \node (species1) at (-1,-5) {
            \resizebox{0.8\columnwidth}{!}{
            \begin{tabular}{ c  c  c  c} 
            \hline  & x-axis & y-axis & z-axis \\ \hline \hline
            Case 1 \ \ & 0.06 / 0.06 \ \ & 0.06 / 0.06 \ \ & 0 / 0 \\ \hline
            Case 2 \ \  & 0.06 / 0.06 \ \ & 0.06 / 0.06 \ \ & -0.5 / +0.5 \\ \hline
            Case 3 \ \ & 0.06 / 0.06 \ \ & 0.06 / 0.06 \ \ & +0.1 / +0.2\\ \hline
            \end{tabular}}
            };
        \node[inner sep=0pt] (russell) at (-5.3,3.25)
            {\normalsize{\textbf{a}}};
        \node[inner sep=0pt] (russell) at (-0.7,3.25)
            {\normalsize{\textbf{b}}};
        \node[inner sep=0pt] (russell) at (-5.3,-0.45)
            {\normalsize{\textbf{c}}};
        \node[inner sep=0pt] (russell) at (-0.7,-0.45)
            {\normalsize{\textbf{d}}};
        \node[inner sep=0pt] (russell) at (-5.3,-4.2)
            {\normalsize{\textbf{e}}};
    \end{tikzpicture}
    \caption{a) Position of the point charges (blue and orange) relative to the molecule to optimize (black dots). 
    Shown is the case of the $\rm{H_2}$ molecule. 
    The orange charges along z-axis are allowed to change (values reported in panel e)) whereas the blue charges are kept fix at the value of $0.06$. 
     b-d) Contour plots of the potential energy generated by the 6 point charges in the xz plane for the three different cases given in the table (panel e). The potentials are in atomic units. 
     e) Value of the charges (in the unit of the fundamental electronic charge $e$) used to generate the 3 different external potentials. 
     For all molecules (A$_1$-A$_2$) in Figure~\ref{fig:fig_2} (panel e) atom A$_1$ has negative $z$-coordinate and A$_2$ a positive one.}
    \label{fig:Fig_System_setup}
\end{figure}

\begin{figure*}[t]
    \centering
        \begin{tikzpicture}
        \node[inner sep=0pt] (russell) at (8.6,-1.1)
            {\includegraphics[width=0.62\textwidth]{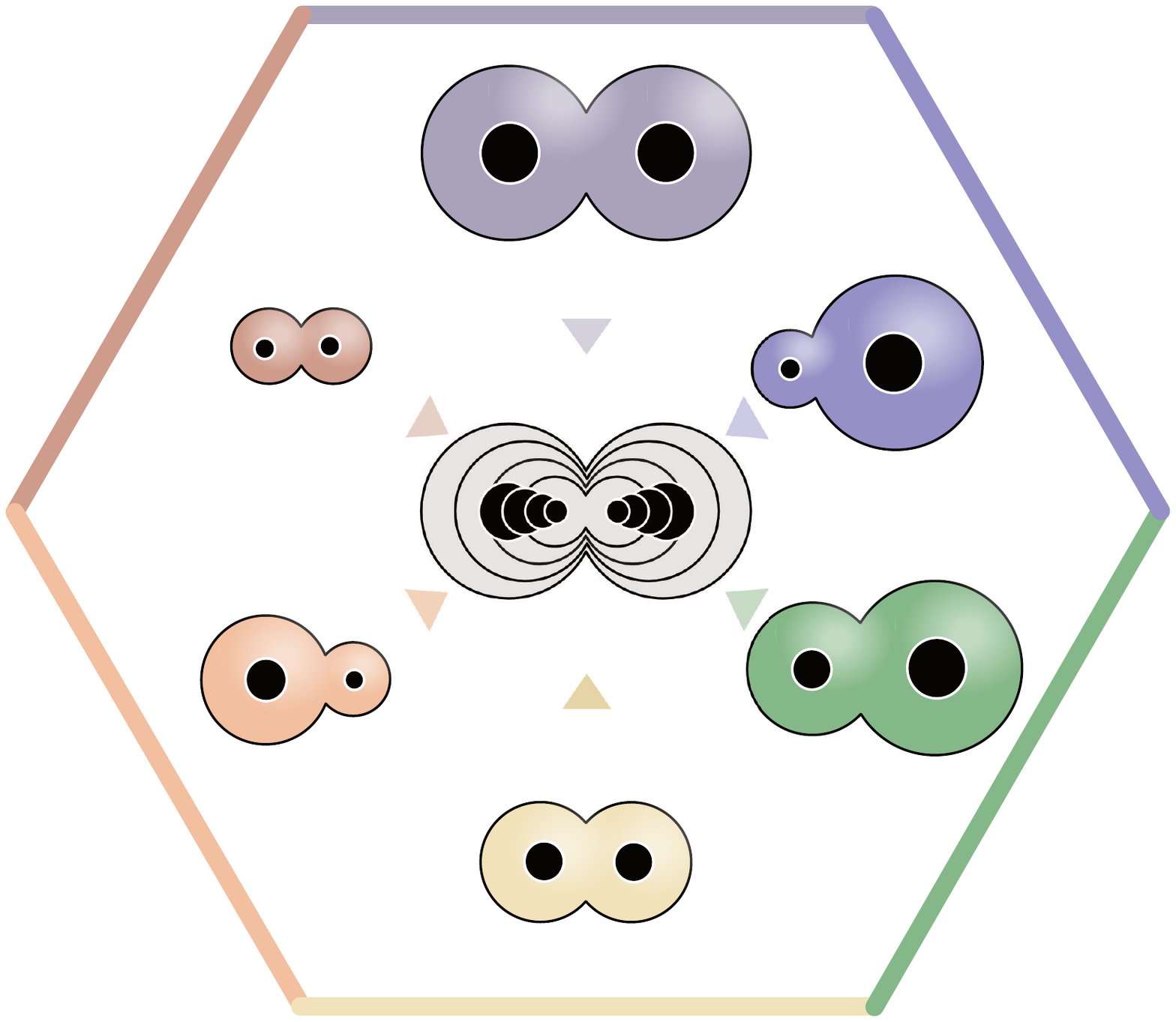}};

        \node at (10.2,-3.5) {\Large{$\dots$}};   
        \node at (9.05,-3.82) {$\Psi_N\left( \theta \right)$};
        \node at (9.1,-4.5) {\Large{$\tilde \alpha_N$}};
        \node at (10.6,-2.9) {$\Psi_5\left( \theta \right)$};
        \node at (11.8,-2.7) {\Large{$\tilde \alpha_5$}};
        \node at (10.6,-1.2) {$\Psi_4\left( \theta \right)$};
        \node at (11.8,0.25) {\Large{$\tilde \alpha_4$}};
        \node at (9.1,0) {$\Psi_3\left( \theta \right)$};
        \node at (9.1,1.8) {\Large{$\tilde \alpha_3$}};
        \node at (7.45,-0.92) {$\Psi_2\left( \theta \right)$};
        \node at (6.2,0.25) {\Large{$\tilde \alpha_2$}};
        \node at (7.45,-2.87) {$\Psi_1\left( \theta \right)$};
        \node at (6.2,-2.7) {\Large{$\tilde \alpha_1$}};
        
        \node at (-1.7,2.2) {D-times};
        
        \node at (3.2,0.45) { \Large{=}};
        
        \draw[dashed] (-2.9,-1.2) -- (0.35,-1.2);
        \draw[dashed] (-2.9,-1.2) -- (-2.9,2.4);
        \draw[dashed] (0.35,2.4) -- (0.35,-1.2);
        \draw[dashed] (0.35,2.4) -- (-2.9,2.4);
        
        \node[inner sep=0pt] (russell) at (-2.6,0.5)
            {
		    \Qcircuit @C=0.4em @R=.6em {
		    &q_{3}: & \quad& \quad &  \quad   & \ket{0} & \quad & \quad &\qw & \gate{R(\theta_{i,0,q_{3}})} &\qw  & \multigate{3}{U_{\rm{ENT}}}& \qw &\gate{R(\theta_{i,d,q_{3}})}&\qw&\qw &\qw\\
		    &q_{2}: & \quad& \quad & \quad   & \ket{0} & \quad & \quad &\qw & \gate{R(\theta_{i,0,q_{2}})} &\qw & \ghost{U_{\rm{ENT}}}	& \qw &\gate{R(\theta_{i,d,q_{2}})}&\qw&\qw &\qw\\
		    &q_{1}: & \quad& \quad & \quad  & \ket{0} & \quad & \quad &\qw & \gate{R(\theta_{i,0,q_{1}})} &\qw & \ghost{U_{\rm{ENT}}}	&\qw &\gate{R(\theta_{i,d,q_{1}})}&\qw&\qw &\qw\\
		    &q_{0}: & \quad& \quad & \quad  & \ket{0} & \quad & \quad &\qw & \gate{R(\theta_{i,0,q_{0}})} &\qw & \ghost{U_{\rm{ENT}}} & \qw &\gate{R(\theta_{i,d,q_{0}})} &\qw&\qw &\qw \\ }
		    };

        \node[inner sep=0pt] (russell) at (3.2,0.5)
            {
		    \Qcircuit @C=0.6em @R=1.63em {
		    & \multigate{3}{U_{\rm{ENT}}} 	& \qw & \quad &\quad &\qw &\ctrl{1} &\qw &\qw &\qw \\
		    & \ghost{U_{\rm{ENT}}}	& \qw & \quad &\quad &\qw &\targ &\ctrl{1} &\qw &\qw\\
		    & \ghost{U_{\rm{ENT}}}	&\qw & \quad & \quad &\qw & \qw &\targ & \ctrl{1} &\qw\\
		    & \ghost{U_{\rm{ENT}}} & \qw & \quad &\quad &\qw &\qw  &\qw & \targ &\qw\\ }
		    };
		    
	    \draw[->] (1.8,-2.8) -- (3,-2.8);
	    \draw[solid] (2.55,-2.8) -- (2.55,-4.7);
	    \draw[solid] (-1.4,-4.7) -- (2.55,-4.7);
	    \draw[->] (-1.4,-4.7) -- (-1.4,-4);
	    \draw[->] (1.1,-4.7) -- (1.1,-4);
	    \draw[->] (-3.7,-2.8) -- (-3.1,-2.8);
	   
	   \node at (-4.4,-2.8) {$\left\{ \theta_{0},\alpha_{0} \right\}$};
	   \node at (3.6,-2.8) {$\left\{ \alpha_{opt} \right\}$};
	   \node at (-2,-4.4) { $\left\{ \theta_{i+1}\right\}$};
	   \node at (0.5,-4.4) { $\left\{ \alpha_{i+1}\right\}$};
	   
	   \node[inner sep=0pt] (russell) at (-0.5,-2.8)
           {
            \Qcircuit @C=1.2em @R=1em {
            & \qw & \multigate{3}{U\left( \theta_i \right)}     & \qw &\qw & \multigate{3}{\langle H\left(\alpha_i \right) \rangle} \\
            & \qw & \ghost{U\left( \theta_i \right)}    &\qw &\qw &\ghost{\langle H\left(\alpha_i \right) \rangle} \\
            & \qw & \ghost{U\left( \theta_i \right)}    &\qw &\qw &\ghost{\langle H\left(\alpha_i \right) \rangle} \\
            & \qw & \ghost{U\left( \theta_i \right)} &\qw &\qw &\ghost{\langle H\left(\alpha_i \right) \rangle} \\ }

           };
		
        \node[inner sep=0pt] (russell) at (2.8,-7.8)
            {\includegraphics[width=0.95\textwidth]{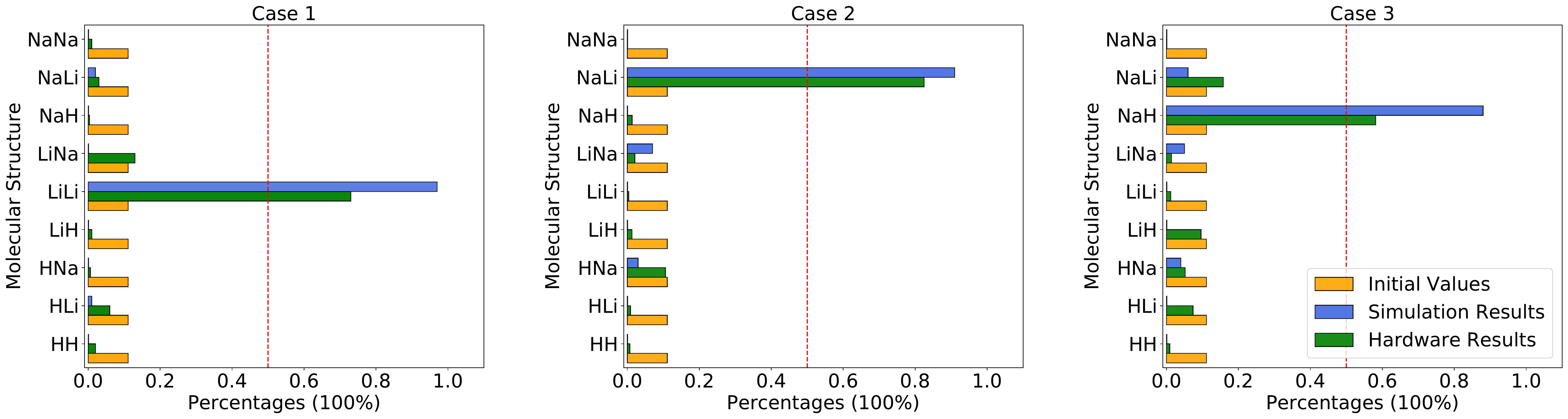}};
    
        \node[inner sep=0pt] (russell) at (-5.9,2.4)
            {\large{\textbf{a}}};
        \node[inner sep=0pt] (russell) at (1,2.4)
            {\large{\textbf{b}}};
        \node[inner sep=0pt] (russell) at (-5.9,-1.6)
            {\large{\textbf{c}}};
        \node[inner sep=0pt] (russell) at (5.4,2.4)
            {\large{\textbf{d}}};
        \node[inner sep=0pt] (russell) at (-5.9,-5.4)
            {\large{\textbf{e}}};
    \end{tikzpicture}
    \caption{
    a) Variational form used to generate the trial alchemical wavefunctions (for this case $D=2$)
    b) Entanglement block used in the variational form for the hardware runs. In classical simulations, we used entangler blocks connecting all qubits. 
    c) Circuit and feedback loop used to simulate and update the wavefunction parameters (circuit variables) and the alchemical weights (of the Hamiltonian). 
    d) Sketch of the molecular superposition state with the contribution of 6 (over $N$) representative molecules participating to the ensemble, each contributing with the probability $\tilde{\alpha}_i=\alpha^1_{S_1(i)} \alpha^2_{S_2(i)}$.
    e) Distribution of the molecular propensities for each of the three charge distributions described in Fig.~\ref{fig:Fig_System_setup}(e).  The initial uniform distribution is shown in orange. The converged VQE distributions obtained in simulations and hardware calculations are given in blue and green, respectively.
    }
    \label{fig:fig_2}
\end{figure*}

\paragraph*{Models and Simulations.}
As a proof-of-principle example, we consider the case of a diatomic molecule placed at the center of the six charges disposed in a bipyramidal arrangement as shown in Fig.~\ref{fig:Fig_System_setup}(a).
The atomic species at each molecular site can be selected from a set composed by the light elements $S=\{\text{H},\text{Li},\text{Na}\}$ of the first column of the periodic table, which are characterized by a single valence electron. The number of possible molecules generated is therefore ${n_s}^{n_p}=3^2$, since due to the potential asymmetry of the axial charges the $XY$ molecule may have a different binding energy than the reversed $YX$ molecule, with $X,Y \in S$.
This choice allows us to keep the number of qubits and the circuit depth of the VQE implementation within the limits of what can be afforded using state-of-the-art simulators and quantum hardware, without limiting the generality of the approach. 
The equatorial charges (as shown in Fig.~\ref{fig:Fig_System_setup}(a) in blue) are all set to the same value, while the axial charges (as shown in Fig.~\ref{fig:Fig_System_setup}(a) in orange) are varied in order to favour different target molecules. The different charge setups are summarized in the Table of Fig.~\ref{fig:Fig_System_setup}(e).

For all three elements we used a STO-3G basis set, which amounts to 1,5,9 atomic basis functions for $\text{H}$, $\text{Li}$, and $\text{Na}$, respectively. 
In order to keep the number of qubits and the number of matrix elements $h_{\mu\nu}(R,\alpha,\tilde{R},q)$ and $g_{\mu\nu\kappa\lambda}(R)$ constant for all molecules, we extended the active space for every molecule according to the minimum active space requirements of the largest molecule.

The bond distances used to compute the matrix elements for the potentials $V_{en}(\tilde{Z}^{s_I}_I,r_i,R_I)$ and $v^{I,s_I}_{\rm{ECP}}(r,R_I)$ of Eq.~\eqref{Hamilt_1st} (which also contribute to $h_{\mu\nu}(R,\alpha,\tilde{R},q)$) are obtained from tabulated values, provided from~\cite{NIST_database}. 
In the case of more complex molecular scaffolds, the algorithm can be extended to include a geometry optimization step where atomic forces are computed from additional measurements of the gradients of the Hamiltonian performed using the `alchemical' wavefunction  $\psi(r;R;\alpha;\tilde{R}; q)$ at each VQE iteration~\cite{o2019calculating}. 
The cost of the geometry optimization step does not modify the scaling of the algorithm but only implies additional $6N_n$ measurements, all performed with the same VQE alchemical wavefunction.

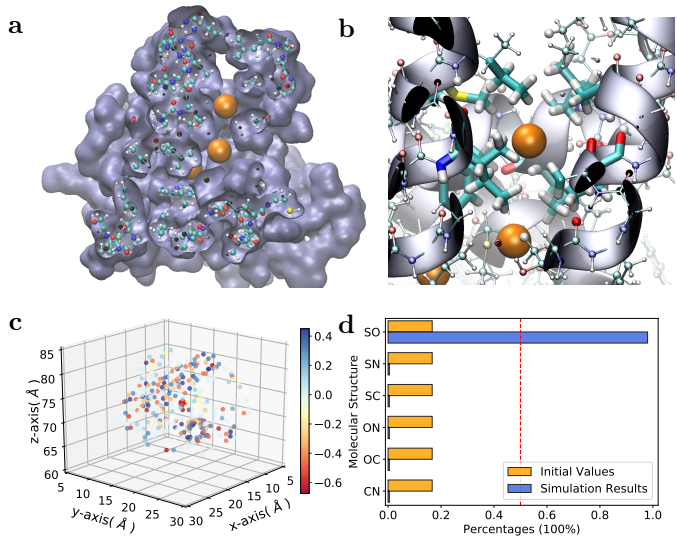
\begin{figure}[h!]
    \centering
    \begin{tikzpicture}
        \node (species1) at (0,0) 
            {\reflectbox{\includegraphics[width=0.95\columnwidth]{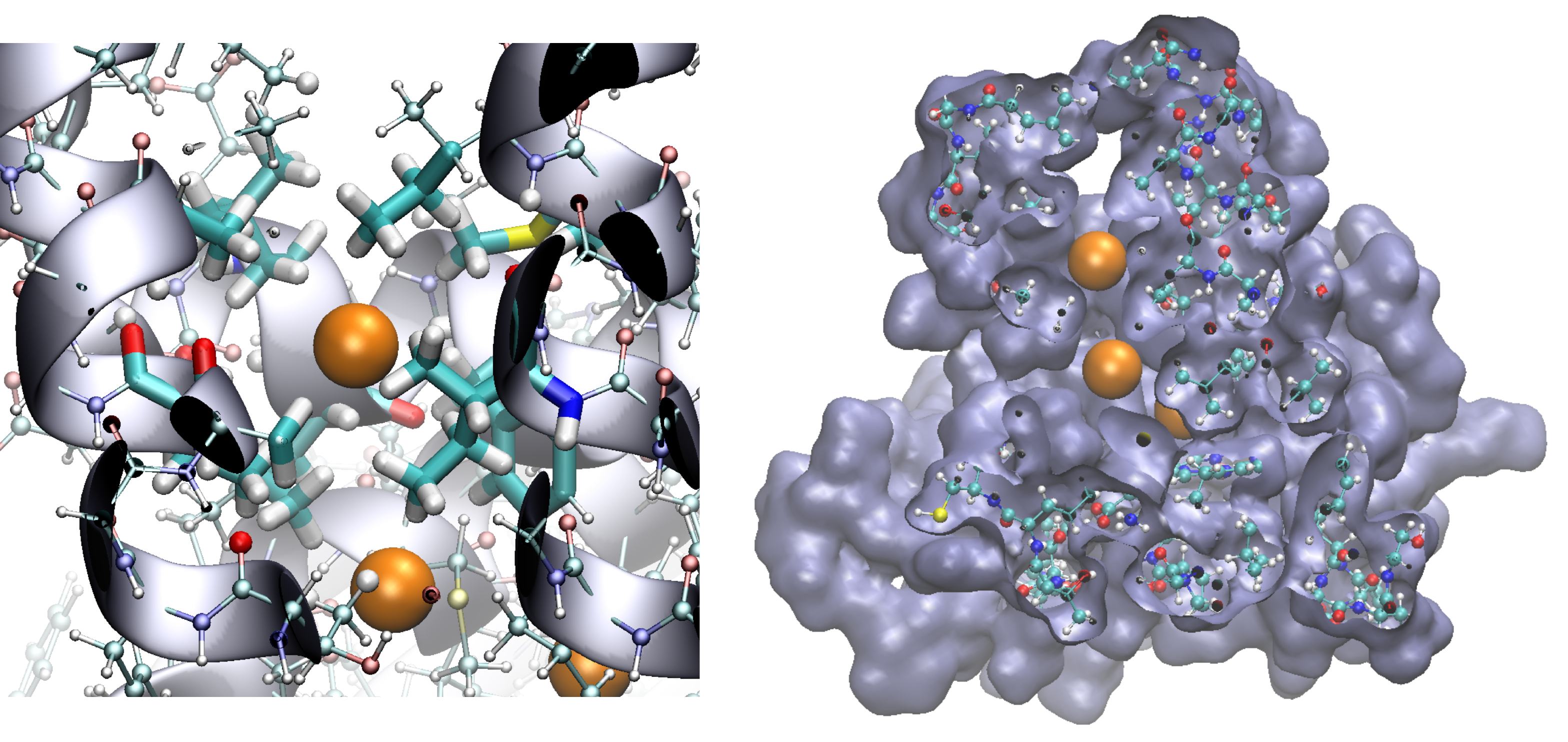}}};
        \node[inner sep=0pt] (russell) at (-0.1,-3.5)
            {\includegraphics[width=0.95\columnwidth]{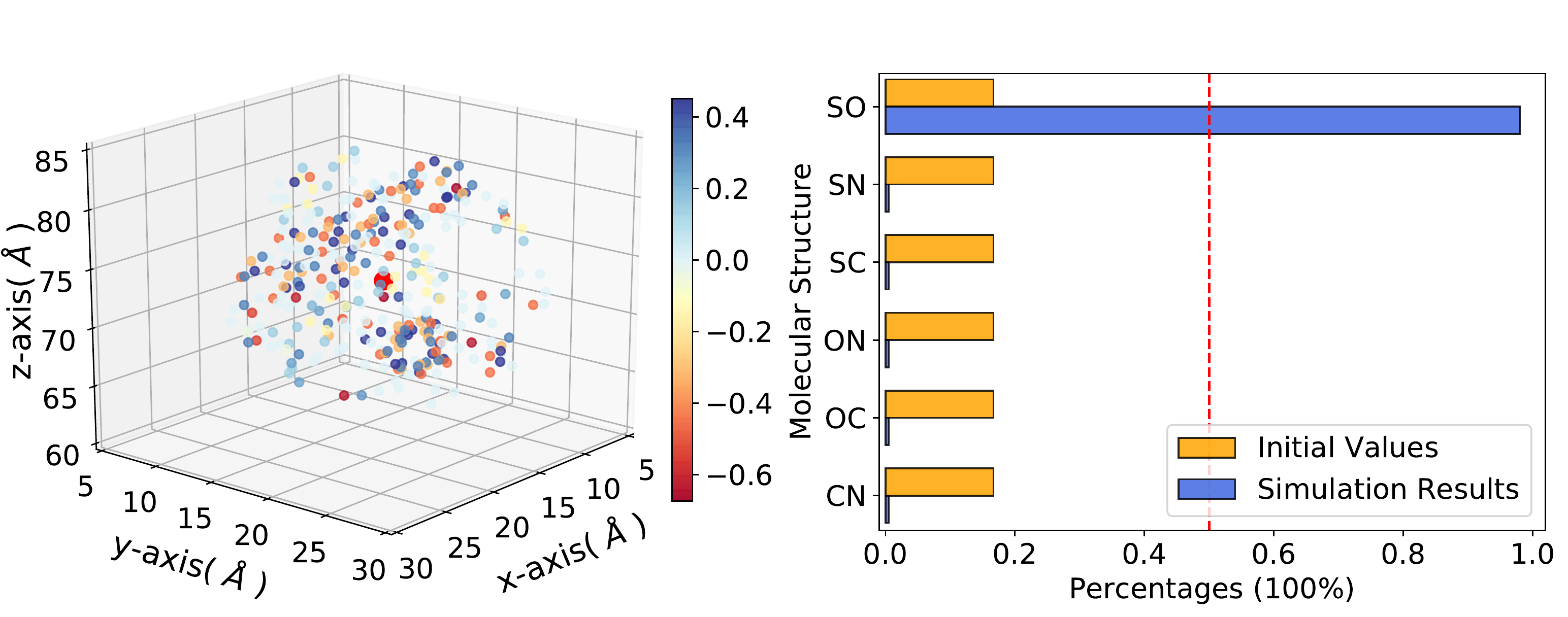}};
        \node[inner sep=0pt] (russell) at (-4.5,1.65)
            {\normalsize{\textbf{a}}};
        \node[inner sep=0pt] (russell) at (-0.1,1.65)
            {\normalsize{\textbf{b}}};
        \node[inner sep=0pt] (russell) at (-4.5,-2.3)
            {\normalsize{\textbf{c}}};
        \node[inner sep=0pt] (russell) at (-0.1,-2.3)
            {\normalsize{\textbf{d}}};
    \end{tikzpicture}
    \caption{a) X-ray crystal structure of the protein H-NOX (PBD entry code 3TFA) showing the main channel with the three main binding pockets occupied by Xe atoms (orange spheres).
    b) Zoom into the first binding pocket with highlighted the apolar residues that favour the selection and diffusion of the gas molecules. The orange spheres indicate the position of the co-crystallized Xe atoms, which mark the center of the binding pockets. c) Position of the point charges used to mimic the enzyme pocket. The red dot corresponds to the center of the molecular structure, d) Initial and final distribution of the molecular structures in the `alchemical' Hamiltonian.}
    \label{fig:protein}
\end{figure}

As a second application, we investigate the binding affinity for different diatomic gases in the binding pocket of the hemeprotein H-NOX (PDB entry code: 3TFA). This protein is involved in sensing and signaling the presence of simple gas molecules in the environment. In particular, in H-NOX one can identify a series of apolar channels that connect the exterior of the protein to the buried heme group. The nature of the hydrophobic pockets arranged along the channels favours the selectivity of the gas molecules and their mobility (see Fig.~\ref{fig:protein}, b). 
In this study, we compute in a single VQE simulation the binding affinity for all possible diatomic molecules that can be generated from the set $\{\text{C},\text{N},\text{O},\text{S}\}$ (the ones chemically unstable are naturally discarded due to their unfavourable formation energies). The center of mass of all molecules is placed at the position of the Xe atom that occupies the binding pocket in the X-ray structure~\cite{Winter2011}. 
The optimal orientation of the diatomic molecules in the pocket is determined for a single element of the ensemble (namely the molecule CO) and kept fix during the VQE optimization.

\paragraph*{Results and discussion.}
\label{Exploring_Various_Configurations}
Fig.~\ref{fig:Fig_System_setup}(e) reports the 3 charge distributions used to test the proposed method.
For each of these charge sets, we optimized the atomic species of the diatomic molecule placed at the center of the charge distribution (see Fig.~\ref{fig:Fig_System_setup}) sampling the elements from the set $\{\text{H},\text{Li},\text{Na}\}$.
Two-dimensional cuts of the electrostatic potentials generated by the different charge arrangements are shown in Fig.~\ref{fig:Fig_System_setup}(b),(c) and (d). 
It is under the influence of these electrostatic potentials that the different atomic compositions for the dimer are selected. 

\renewcommand{\arraystretch}{1.3}

The VQE is initialized using an unbiased distribution of the atomic weights associated to each atomic position, namely $\alpha^I_{s_I}=1/N_I N_{s_I}, \forall s_I \, \forall I$ so that $\sum_{s_I}\alpha^I_{s_I}=1$, $\forall I$. 
All $N_b^t$ qubits are initialized in state  `1' (note that the choice of atomic basis functions in place of molecular (HF) ones implies the possible occupation of all orbitals in the generation of the system wavefunction).
As a consequence, the circuit used to sample the `alchemical' wavefunction is not required to conserve the total occupation number (i.e., the number of `1' in the qubit register).
Fig.~\ref{fig:fig_2}(e) shows the evolution of the atomic compositions at the two molecular sites of the system in Fig.~\ref{fig:fig_2}(a) along the optimization process for different choices of the external charges (see Table in Fig.~\ref{fig:Fig_System_setup}(e)). 
Starting from the initial equally distributed atomic compositions (i.e., $\alpha_{s_I}^I=1/N^I_s$) in all cases we observe a smooth increase of the population of a given atomic species at each of the two atomic positions. 
As expected, the most probable final structure depends on the choice of the external potential. 
The correctness of our predictions are confirmed with an \textit{a posteriori} calculation of the binding energies using an equivalent classical algorithm.

In the case of H-NOX simulation, we determined the diatomic molecule obtained combining two elements from the set $\{\text{C},\text{O},\text{N},\text{S}\}$, which has the best affinity for the first binding pocket exposed to the solvent. The simulation of the `alchemical' quantum algorithm proposed in this work unequivocally selects the molecule SO, independently from the orientation of the molecular axis. The solution is in agreement with the `classical' solution obtained by scanning the binding energy of all possible molecules. 

\paragraph*{\label{sec:M&M}Materials and Methods}

In order to ensure that the `alchemical' Hamiltonian can be encoded with the same number of qubits we use the same number of basis function for every atomic composition. 
This requires the extension of active space towards non-contributing basis functions for the atomic compositions with less valence electrons. 
For the combination of atoms ${\text{C},\text{O},\text{N},\text{S}}$ the minimum space required would be of 8 atomic orbitals (16 qubits) in total for the active space approximation for the $\text{SN}$ atomic combination.

The whole simulation of the `alchemical' Hamiltonian is executed in the Qiskit~\cite{Qiskit} framework. 
For the classical optimization procedure we employ the \textit{SLSQP} optimizer as integrated in Qiskit Aqua. 
To ensure that the alchemical parameters stay within the bounds of 0 and 1 we implemented the corresponding bounds for the \textit{SLSQP} algorithm and a general constraint of the form $\sum_i \alpha_i = 1$.
The initial values for the parameterized wavefunction were set as random values within the $[0,\pi]$ interval for the first run. 
For all subsequent runs we used as initial values the wavefunction parameters that were estimated in the first run.
The bounds for the $\theta$ parameters within the optimization process where set to be from $0$ to $2\pi$. 
For the classical simulations runs we set the maximum number of iterations to $500$, while for real hardware we used $100$ iterations. 

In the optimization process we observed that by multiplying the energy gap  by a factor $f$ allows for a faster convergence of the VQE. For this reason, the minimization was driven by cost function ${\rm{min}}_{\theta} \braket{ f \left(H_C(R,\alpha) - H(R,\alpha) \right)}_{\ket{\psi\left(\theta \right)}}$ while the binding energies are calculated using $(\braket{f \left(H_C(R,\alpha) - H(R,\alpha) \right)}_{\ket{\psi\left(\theta \right)}})/f$.
For the results in Fig.~\ref{fig:fig_2} a factor $f=10^3$ was used while for Fig.~\ref{fig:protein} we chose $f=10^4$. 
To improve the convergence of the algorithm, we performed several restarts of the VQE using as initial parameters the ones outputted in the previous optimization loop. 

The hardware runs were executed on \textit{ibmq\_singapore} 20-qubit chip via Qiskit~\cite{Qiskit}.
For all runs we used 8192 shots and we used the natural connectivity of the qubits to make the variational form (Fig~\ref{fig:fig_2}(a) and Fig~\ref{fig:fig_2}(b)).

\paragraph*{Conclusions.}
In this work, we present a method for the design of molecular systems that best fit with a given external potential using an `alchemical' quantum algorithm that simultaneously optimizes the electronic structure and the nuclear composition of given compound.
The search occurs in the exponentially large chemical compound space, which hampers an efficient implementation using classical algorithms, also due to the cost associated to the solution of the SE. 
The advantage of our approach is two-fold: 
it leverages the favourable scaling of quantum electronic structure algorithms ($\mathcal{O}(N^4)$ in the number of basis functions) as well as the possibility to search in an exponentially large space using a polynomial number of resources (i.e., number of qubits and gate operations).

The algorithm was successfully applied for the optimization of a diatomic molecule composed by elements from the ensemble $\{\text{H},\text{Li},\text{Na}\}$ placed in an external potential generated by 6 point charges. Simulations and hardware calculations performed on \textit{ibmq\_singapore} chip could unequivocally identify the best fitting molecule over an ensemble of 9 possible structures. 
To further validate our approach, we also apply the same algorithm for the determination of the diatomic molecules (with elements from the ensemble $\text{C},\text{O},\text{N}$ and $\text{S}$) that best fit in the binding pocket of the hemeprotein H-NOX represented by 330 charges. Also in this case, the simulation correctly selects the most stable molecule, namely SO as confirmed from electronic structure and DFT calculations.

In conclusion, we propose and demonstrate a quantum algorithm which enables the optimization of chemical structures in a given external potential.
The results illustrate the potential of quantum algorithms as a tool for the efficient and accurate sampling of the chemical compound space and open up new perspectives for the use of quantum computers in material design and drug discovery applications.

\section*{Acknowledgements}
P.J.O, I.O.S and I.T. acknowledge financial support from the Swiss National Science Foundation (SNF) through the grant No. 200021-179312.
All authors acknowledge support from Ismael Faro and interesting discussions with Alessandro Curioni and the whole IBM Quantum team. 

IBM, IBM Q, IBM Quantum, Qiskit are trademarks of International Business Machines Corporation, registered in many jurisdictions worldwide. Other product or service names may be trademarks or service marks of IBM or other companies.

\bibliography{alchemical_lib}
\end{document}